\documentstyle[prl,aps,multicol]{revtex}

\begin{document}

\title{Nonlinear Spectral Characterization of Discrete Data}
\author{M. BOITI$^{*}$, J. LEON, F. PEMPINELLI$^{*}$ \\
Physique Math\'ematique et Th\'eorique, CNRS, F-34095 MONTPELLIER\\
$^{*}$ Dipartimento di Fisica, Universit\`{a} di Lecce and Sezione INFN,
I-73100 LECCE.}
\maketitle

\begin{abstract}
The {\em explicit analytical expression} of the Nonlinear Fourier Transform
(NFT) of a finite set of data is provided. Then a simple {\em recursion
relation} for the NFT is constructed as a {\em function} of the spectral
parameter. These tools provide a  {\em complete characterization} of 
the nonlinear coherent structures (solitons, breathers, ...) 
present in numerical or experimental data representing the
solution, at a given value of time, of a nonlinear evolution equation (e.g.
of the nonlinear Schr\"odinger family).
\end{abstract}

\draft


\begin{multicols}{2}

\paragraph*{Introduction.}

The inverse scattering (or spectral) transform (IST) is a method for
constructing and solving nonlinear evolution equations (NEE) \cite{bookist}.
One of those, the nonlinear Schr\"odinger equation (NLS) \cite{zakhshab72},
has become a paradigm as being the model for a wide variety of physical
situations. Actually the relevant equation generally turns out to be {\em %
nonintegrable}, e.g. NLS with forcing and damping, which renders necessary a
numerical analysis \cite{makhankov83}. Then, a fundamental question is the
characterization of the solution, as a standard Fourier analysis will not
individualize the intrinsically nonlinear structures (as solitons). The same
question is even more essential when dealing with experimental data.

Another fundamental aspect  is the question
of the {\em discretization of an integrable continuous model}, which rises
many interesting questions such as numerical induced chaos \cite{ablow-89},
instabilities in integrable systems \cite{forest-92} or roundoff error
growth \cite{ablow-93}.

Moreover, localized excitations (solitons) in nonlinear chains has been a
subject of intense research for their obvious physical interest \cite
{books-discr}\cite{toda}\cite{siev-tak}. It has been proved for instance
that the discreteness generically helps energy localization \cite{peyra-93}
\cite{flach-94}, that it is a mechanism controlling wave collapse \cite
{bishop-94}, and also that it allows for the existence of moving localized
modes \cite{claude-93} or standing ones \cite{kivshar-94}.  

In all these studies, the existence and
properties of such localized excitations result merely from a direct {\em %
observation} of the numerical solution of the model. In short a soliton
is recognized from its shape, its velocity from the velocity of the point
of maximum amplitude, the frequency of a breather is obtained from its Fourier
spectrum, etc...

However there exists a tool, the {\em nonlinear Fourier transform} (NFT),
which allows to characterize completely the nonlinear modes. It is based on
IST and has been developed numerically either for the sine-Gordon family
\cite{overman} to modelize fluxon dynamics in Josephson junctions, or for the
Korteveg-de Vries model \cite{osborne} to modelize water waves. In that case,
the method has been successfully applied to experimental data
\cite{osborne-2}, for instance to prove that solitons {\em generically} occur
in wave trains. The drawback however is on the one side the complexity of the
method, and on the other side the lack of analytical formula for the NFT.

This letter is intended to provide two results, essential for the 
{\em nonlinear analysis} of discrete data. The first one is an {\em analytical
explicit formula} (\ref{alfexpl}) for the NFT of a discrete set of data.
More precisely, taking advantage of the finiteness of numerical or
experimental data, we obtain the explicit expression of the Jost solutions,
and hence of the complete spectrum characterization, corresponding to a
{\em discrete finite potential}. The second one is a quite simple recursion
relation (\ref{alpha-recursion}) for a very straightforward numerical
construction of the spectrum.

Everything is done in the context of the Ablowitz-Ladik spectral problem
\cite{abladi} (discrete version of the Zakharov-Shabat spectral problem),
which means that the nonlinear Fourier analysis applies here for instance to
the NLS family, that is
\begin{eqnarray}
i\partial_tq(n)&=&q(n{+}1)+q(n{-}1)-2q(n)\pm  \nonumber \\
&&|q(n)|^2[q(n{+}1)+ q(n{-}1)]+a(n)  \label{NLS}
\end{eqnarray}
where $a(n)$ is any perturbation term. The method works without change for 
all nonintegrable deformations of the others integrable NEE related to the
Ablowitz-Ladik spectral problem, like mKdV, Toda, self-dual network, KdV,
sine-Gordon \cite{ablow-seg}, or discrete Stimulated Raman Scattering \cite
{papier}. Last, the tool can be successfully applied to experimental data,
as soon as the measurements are assumed to represent a train of envelope
wave pulses which can be considered to vanish outside a finite interval.
Indeed in such a case the model equation for the envelope is likely to obey
a member of the NLS family.


\paragraph*{Statement of the problem.}

Given a finite discrete set of $2(L+1)$ complex valued data $q(n),r(n)$
having significant values on $n=0,\dots,L$ and vanishing outside, the
problem is to determine its nonlinear Fourier spectrum (the scalar case will
enter as a reduction, e.g. $r=\mp\bar q$ to recover (\ref{NLS})). This is
done by using these data as the {\em potential} of the Ablowitz-Ladik
scattering (or spectral) problem:
\begin{eqnarray}
&&\mu _{11}(n{+}1)-\mu _{11}(n)=q(n{+}1)\mu _{21}(n{+}1),  \nonumber \\
&&\mu _{21}(n{+}1)-k^{-1}\mu _{21}(n)=r(n{+}1)\mu _{11}(n{+}1),  \nonumber \\
&&\mu _{12}(n{+}1)-k\mu _{12}(n)=q(n{+}1)\mu _{22}(n{+}1),  \nonumber \\
&&\mu _{22}(n{+}1)-\mu _{22}(n)=r(n{+}1)\mu _{12}(n{+}1),
\label{Ablw-Lad-spec}
\end{eqnarray}
for $n=-1,\ldots ,L-1$. A dependence on the {\em spectral parameter} $k$ is
understood everywhere for the matrix valued function $\mu(k,n)$ with
elements $\mu_{ij}$.

The Jost solutions $\mu ^{\pm }$ are then defined by the following discrete
integral equations
\begin{mathletters}
\label{mu-jost}
\begin{eqnarray}
&&\left(\matrix{\mu _{11}^{-}(n) \cr \mu _{21}^{-}(n)}\right) = \left(%
\matrix{1 \cr 0}\right) +\left( \matrix{
-\displaystyle\sum_{i=n+{1}}^Lq(i)\mu _{21}^{-}(i) \cr
\displaystyle\sum_{i=0}^nk^{i-n}r(i)\mu _{11}^{-}(i)} \right),
\label{mu1-moins} \\
&&\left( \matrix{ \mu _{11}^{+}(n) \cr \mu _{21}^{+}(n)}\right) = \left( %
\matrix{1 \cr 0}\right) - \left( \matrix{
\displaystyle\sum_{i=n+{1}}^Lq(i)\mu _{21}^{+}(i) \cr
\displaystyle\sum_{i=n+{1}}^Lk^{i-n}r(i)\mu _{11}^{+}(i)} \right),
\label{mu1-plus} \\
&&\left( \matrix{ \mu _{12}^{-}(n) \cr \mu _{22}^{-}(n)} \right) =\left( %
\matrix{0 \cr 1} \right) - \left( \matrix{
\displaystyle\sum_{i=n+{1}}^Lk^{n-i}q(i)\mu _{22}^{-}(i) \cr
\displaystyle\sum_{i=n+{1}}^Lr(i)\mu _{12}^{-}(i) }\right),
\label{mu2-moins} \\
&&\left( \matrix{ \mu _{12}^{+}(n) \cr \mu _{22}^{+}(n) }\right) =\left( %
\matrix{0 \cr 1}\right)+ \left( \matrix{
\displaystyle\sum_{i=0}^nk^{n-i}q(i)\mu _{22}^{+}(i) \cr
-\displaystyle\sum_{i=n+{1}}^Lr(i)\mu _{12}^{+}(i)} \right).
\label{mu2-plus}
\end{eqnarray}


\paragraph*{Nonlinear Fourier spectrum.}

The spectral transform of $\{q,r\}$ is then completely determined by the
Jost solutions $\mu^+$ and $\mu^-$ as follows. An essential function is the
{\em reflection coefficient} $\alpha^\pm(k)$ defined as
\end{mathletters}
\begin{equation}
\alpha ^{-}=\sum_{0}^{L}k^ir(i)\mu _{11}^{-}(i), \quad\alpha
^{+}=\sum_{0}^{L}k^{-i}q(i)\mu _{22}^{+}(i).  \label{alpha-def}
\end{equation}
Due to the finiteness of the support of the potential $\{q,r\}$, the Jost
eigenfunction, and hence $\alpha^\pm(k)$, can be defined in the whole
complex $k$-plane. Then the knowledge of the reflection coefficient in the
complex plane allows to determine completely the nonlinear Fourier spectrum,
which consists of a continuous part ({\em radiation}) related to the values
of $\alpha^\pm$ on the unit circle $|k|=1$, and of a discrete part ({\em %
solitons}) constituted by the poles $k_j^-$ of $\alpha^-$ in $|k|>1$ and $%
k_j^+$ of $\alpha^+$ in $|k|<1$. It is our purpose here to give an explicit
formula for $\alpha^\pm$ in terms of the complex variable $k$.


\paragraph*{Explicit solution.}

Introducing, for $(k,n)$ the new functions
\begin{equation}
\left(\matrix{\nu^\pm_{11}\cr\nu^\pm_{21}}\right)= \left(\matrix{\mu^%
\pm_{11}\cr k^n\mu^\pm_{21}}\right),\quad \left(\matrix{\nu^\pm_{12}\cr\nu^%
\pm_{22}}\right)= \left(\matrix{k^{-n}\mu^\pm_{12}\cr \mu^\pm_{22}}\right),
\label{new-nu}
\end{equation}
the integral equations (\ref{mu-jost}) become explicitly solvable in terms
of the new matrix $\nu(k,n)$. This solution can be written in $(k,n)$
\begin{equation}
\left(\matrix{\nu^-_{11}\cr\nu^-_{21}}\right)= X\ \left(\matrix{1\cr\alpha^-}%
\right),\quad \left(\matrix{\nu^+_{11}\cr\nu^+_{21}}\right)= X\ \left(%
\matrix{1\cr0}\right),  \label{nu-1}
\end{equation}
\begin{equation}
\left(\matrix{\nu^-_{12}\cr\nu^-_{22}}\right)= X\ \left(\matrix{0\cr1}%
\right),\quad \left(\matrix{\nu^+_{12}\cr\nu^+_{22}}\right)= X\ \left(%
\matrix{\alpha^+\cr1}\right).  \label{nu-2}
\end{equation}
The matrix $X(n)$ reads
\begin{equation}
X(n)=\prod_{i=n+1}^L(1-A_i), \quad A_i = \left(\matrix{0 & k^{-i}q(i) \cr
k^ir(i) & 0}\right)  \label{X-1}
\end{equation}
with $X(L)=1$, or else
\begin{equation}
X(n)=1-\sum_{\text{odd}\;l=1}^{L-n} \left(\matrix{0 & Q_l \cr R_l & 0}%
\right)+ \sum_{\text{even}\;l=2}^{L-n} \left(\matrix{ U_l & 0 \cr 0 & V_l}%
\right)  \label{X-2}
\end{equation}
with the following definitions of the functions of the two variables 
$(l,n)$ :
\begin{eqnarray}
&&Q_l=\sum k^{-j_1}q(j_1)k^{j_2}r(j_2)\ldots k^{-j_l}q(j_l),  \nonumber \\
&&R_l=\sum k^{j_1}r(j_1)k^{-j_2}q(j_2)\ldots k^{j_l}r(j_l)  \nonumber \\
&&U_l=\sum k^{-j_1}q(j_1)k^{j_2}r(j_2)\ldots k^{j_l}r(j_l),  \nonumber \\
&&V_l=\sum k^{j_1}r(j_1)k^{-j_2}q(j_2)\ldots k^{-j_l}q(j_l).
\end{eqnarray}
The sums run on all possible different $l$ indices $j_i$, ordered for each
value of $(l,n)$ as ${n+1\leq j_1<j_2<\ldots <j_l\leq L}$.

Then we may compute from $X(n)$ the reflection coefficients $\alpha^\pm(k)$
which from their definition (\ref{alpha-def}) read
\begin{eqnarray}
\alpha ^{-}(k)&=&\frac{\sum_{i=0}^Lk^ir(i)X_{11}(i)}{1-%
\sum_{i=0}^Lk^ir(i)X_{12}(i)}\ ,  \nonumber \\
\alpha ^{+}(k)&=&\frac{\sum_{i=0}^Lk^{-i}q(i)X_{22}(i)}{1-%
\sum_{i=0}^Lk^{-i}q(i)X_{21}(i)}\ .  \label{alfexpl}
\end{eqnarray}
Note: the reduction $r(n)=\pm\bar q(n)$ leading e.g. to the NEE (\ref{NLS}),
has the counterpart \cite{papier} $\alpha^+(1/k)=\pm \overline{\alpha^-}%
(\bar k)$. These two relations are indeed compatible with (\ref{alfexpl}).


\paragraph*{Data reconstruction.}

The data $\{r,q\}$ can be reconstructed from their spectral transform $%
\{\alpha^+,\alpha^-\}$ because the solutions $\mu^\pm(k,n)$ of (\ref{mu-jost}%
) also solve \cite{papier}
\begin{mathletters}
\label{mu-cauchy}
\begin{eqnarray}
&&\left(\matrix{\mu_{11}^\pm(k)\cr\mu_{21}^\pm(k)}\right) = \left(\matrix{1
\cr 0}\right)- \frac{1}{2\pi i}\oint\frac{\zeta^{-n}d\zeta\alpha^-(\zeta) }{%
\zeta-(1\mp 0)k}\ \frac{k}{\zeta} \left(\matrix{\mu_{12}^{-}(\zeta)\cr%
\mu_{22}^{-}(\zeta)}\right)  \nonumber \\
&&+\sum_j(k_j^-)^{-n} \displaystyle{\mathop{\text Res}_{k_j^-}}\{\alpha^-\}
\frac{k}{k_j^-}\frac{1}{k_j^--k} \left(\matrix{\mu_{12}^{-}(k_j^-)\cr%
\mu_{22}^{-}(k_j^-)}\right),  \label{mu1-cauchy} \\
&&\left(\matrix{\mu_{12}^\pm(k)\cr\mu_{22}^\pm(k)}\right)= \left(\matrix{0
\cr 1}\right)+ \frac{1}{2\pi i}\oint\frac{\zeta^{n}d\zeta\alpha^+(\zeta) }{%
\zeta-(1\mp 0)k}\ \left(\matrix{\mu_{11}^{+}(\zeta) \cr\mu_{21}^{+}(\zeta)}%
\right)  \nonumber \\
&&-\sum_j(k_j^+)^{n} \displaystyle{\mathop{\text Res}_{k_j^+}}\{\alpha^+\}
\frac{1}{k_j^+-k} \left(\matrix{\mu_{11}^{+}(k_j^+)\cr\mu_{21}^{+}(k_j^+)}%
\right),  \label{mu2-cauchy}
\end{eqnarray}
where the integral on $\zeta$ runs on the unit circle.

The system (\ref{Ablw-Lad-spec}) at order zero in $k$ then gives
\end{mathletters}
\begin{equation}
q(n{+}1)=-\mu _{12}^-(n)^{(-1)},\quad r(n{+}1)=-\mu _{21}^+(n)^{(1)},
\label{potentials}
\end{equation}
where $\mu _{12}^-(n)^{(-1)}$ is the coefficient of $k^{-1}$ in the Laurent
expansion for $k\to \infty $ of $\mu _{12}^{-}(k,n)$, and $%
\mu_{21}^+(n)^{(1)}$ the coefficient of $k$ in the Taylor expansion for $%
k\to 0$ of $\mu _{21}^{+}(k,n)$.

These formulae are useful to prove the {\em consistency} of the method, that
is that the reconstructed data are the same as the starting ones (in
particular that they stay in the class of finite support). We do not develop
this lengthy proof here but rather make the following remark: the above
formulae allows one to {\em filter} the data just by setting to zero any
part of the spectrum. For instance, if one needs information on the
underlying nonlinear coherent structures, one may set to zero the
contribution $\alpha^\pm(\zeta),\ |\zeta|=1$ to the radiation, and then
solve (\ref{mu-cauchy}). Of course, doing this the reconstructed data do not
stay in the class of {\em finite support potentials}, and this procedure
will be meaningful for potentials going to zero fast enough at both ends.


\paragraph*{Recursion relation.}

From its very definition (\ref{X-2}), the matrix $X(n)$ can actually be
considered as a function also of the {\em variable} $L$ (remember that $L$
is the dimension of the data support). We shall then denote it by $X^L(n)$
and for instance $X^m(n)$ has to be understood as obtained from the data of $%
\{q(n),r(n)\}$ for $n=0,1,\dots, m$, and zero for $n \notin [0,m]$. Then
our purpose here is to derive a recursion relation for the matrix elements
of $X^m(n)$, in order to obtain a recursion relation for the corresponding
spectral data $\alpha^\pm_m(k)$, spectral transform of $\{q(n),r(n)\}$ for $%
n=0,1,\dots, m$.

The spectral transform (\ref{alfexpl}) is first conveniently rewritten as
\begin{equation}
\alpha_m^{-} =\frac{S_m}{1+\sum_{j=0}^{m-1}p_{j+1}S_j}, \quad \alpha_m^{+} =%
\frac{P_m}{1+\sum_{j=0}^{m-1}s_{j+1}P_j},  \label{alphaexpl-2}
\end{equation}
with the following definitions
\begin{eqnarray}
S(m)=\sum_{n=0}^ms(n)X_{11}^m(n), && P(m)=\sum_{i=0}^mp(n)X_{22}^m(n),
\nonumber \\
p(k,n)=k^{-n}q(n), && s(k,n)=k^nr(n).  \label{p-s-def}
\end{eqnarray}
Then only the diagonal elements of $X^m(n)$ have to be considered.

By a careful rewriting of the element $X_{11}^{m+1}(n)$ out of (\ref{X-2}),
and by using that obviously $X_{11}^{n}(n)=X_{11}^{n+1}(n)=1$, we obtain
\begin{equation}
X_{11}^{m+1}(n)=X_{11}^m(n)+s(m+1)\sum_{j=n}^{m-1}p(j+1)X_{11}^j(n).
\label{X-11}
\end{equation}

A similar recursion relation is immediately deduced for $S(m)$ defined
above, and consequently for $\alpha_m^{-}$. The same computation being made
for $X_{22}^{m+1}(n)$ leads finally to
\begin{eqnarray}
\alpha _0^{-} =r(0), \quad && \alpha _{m+1}^{-} =\frac{s_{m+1}+\alpha _m^{-}%
}{1+p_{m+1}\alpha _m^{-}},  \nonumber \\
\alpha_0^{+} =q(0), \quad && \alpha_{m+1}^{+}=\frac{p_{m+1}+\alpha _m^{+}}{%
1+s_{m+1}\alpha _m^{+}}.  \label{alpha-recursion}
\end{eqnarray}
If the data to be analyzed are given on $[0,L]$, the spectral transform $%
\alpha^\pm(k)$ is constructed from the above recursion relation by running $%
m $ from 0 to $L$.
To illustrate the preceding results we consider two examples.

\paragraph*{1$^{\mbox{st}}$ Example: two-point data.}

In the case of two-point information ($L=1$), by direct
application of the above formula, we get
\begin{equation}
\alpha^{-}_1=\frac{r(0)+kr(1)}{1+k^{-1}q(1)r(0)}\ ,\quad \alpha^{+}_1=\frac{%
q(0)+k^{-1}q(1)}{1+kq(0)r(1)}\ .
\end{equation}

An obvious and natural consequence is that two point data {\em cannot}
represent a pure soliton solution which would require that $\alpha^\pm$
vanish on the unit circle $|k|=1$. The important information is that one can
readily conclude about the presence of a soliton if $|q(1)r(0)|>1$ (which
ensures a pole of $\alpha^-$ in $|k|>1$) or/and if $|q(0)r(1)|>1$ (which
ensures a pole of $\alpha^+$ in $|k|<1$).

\paragraph*{2$^{\mbox{nd}}$ Example: truncated soliton.}

We consider the one soliton solution (reduction $r=-\overline{q}$) centered
on $m=n_0$
and cut to the left of $m=0$ and to the right of $m=L$ defined by ($\eta_0$, 
$\theta _0$ arbitrary constants and $b>0$)
\begin{equation}
q(m)=\frac{e^{i\eta _0m+\theta _0}}{\tau_m},\quad \quad \tau_m=\frac{%
\cosh \left( bm-bn_0\right) }{\sinh b}
\end{equation}
for $m=0,1,\cdots ,L$ and $q(m)=0$ for $m<0$ and $m>L.$ By using the
recursion formula (\ref{alpha-recursion}) the spectral transform can be
explicitly computed giving ($\xi =ke^{-i\eta _0}$)
\begin{equation}
\alpha_L^+=-e^{i\theta _0}\xi ^{-L}\frac{\xi ^{L+2}\tau_L-\xi ^{L+1}\tau_{L+1}
-(\xi \tau_{-1}-\tau_0)}{\xi ^{L+2}-(\xi \tau_{L+1}-\tau_L)(\xi
\tau_{-1}-\tau_0)}.
\end{equation}
A simple analysis for $L\gg n_0$ shows that $\alpha _L$ has a pole for $|k|<1$
if and only if $n_0>0$. This means that the presence of a soliton
in a wave train is detected with this method
by analyzing only a {\em finite portion} of it. In
the example considered, it is in fact enough to catch half of the soliton.

\paragraph*{Conclusion.}

The method is quite simple to use and it is of wide application: the
formulae (\ref{alfexpl}) provide an explicit function of the complex
variable $k$ which allows to completely characterize the nonlinear Fourier
spectrum of the data $\{r(n),q(n)\}$. In particular the roots of the
denominator of $\alpha^-$ in the region $|k|>1$, and those of $\alpha^+$ in
$|k|<1$, furnish the nonlinear coherent structures present in the data. For
practical applications, the recursion relations (\ref{alpha-recursion})
provide a quite efficient numerical code to generate the nonlinear Fourier
spectrum of any finite data. These relations possess a continuous
counterpart which will be studied, together with details and applications,
in a forthcoming paper.

\paragraph*{Aknowledgements.}

MB and FP are grateful to the {\em Laboratoire de Physique Math\'ematique et
Th\'eorique} for hospitality and to the CNRS for support. This work is part
of the contract {\em Dynamique nonlin\'eaire et chaines mol\'eculaires - UM2}%
.

\end{multicols}

\end{document}